\documentclass[prl,aps,
showpacs,
preprint,
nofootinbib,
floatfix,
superscriptaddress, showkeys
]{revtex4-1}
\usepackage{graphics}
\usepackage{epsfig}
\usepackage{latexsym}
\usepackage{colordvi}
\usepackage{amsmath}
\usepackage{amssymb}
\newcommand{\be}{\begin{equation}}
\newcommand{\ee}{\end{equation}}
\newcommand{\bea}{\begin{eqnarray}}
\newcommand{\eea}{\end{eqnarray}}

\def\IB{\relax\hbox{$\inbar\kern-.3em{\rm B}$}}
\def\IC{\relax\hbox{$\inbar\kern-.3em{\rm C}$}}
\def\ID{\relax\hbox{$\inbar\kern-.3em{\rm D}$}}
\def\IE{\relax\hbox{$\inbar\kern-.3em{\rm E}$}}
\def\IF{\relax\hbox{$\inbar\kern-.3em{\rm F}$}}
\def\IG{\relax\hbox{$\inbar\kern-.3em{\rm G}$}}
\def\IGa{\relax\hbox{${\rm I}\kern-.18em\Gamma$}}
\def\IH{\relax{\rm I\kern-.18em H}}
\def\IK{\relax{\rm I\kern-.18em K}}
\def\IL{\relax{\rm I\kern-.18em L}}
\def\IP{\relax{\rm I\kern-.18em P}}
\def\IR{\relax{\rm I\kern-.18em R}}
\def\IZ{\relax{\rm Z\kern-.5em Z}}

\begin{document}
\preprint{\parbox[b]{1in}{ \hbox{\tt PNUTP-17/A04}}}
%\hfill\parbox[b]{1in}{ \hbox{\tt PNUTP-17/A04}}
\title{Casimir scaling and Yang-Mills glueballs}

\author{Deog Ki Hong}
\email[E-mail: ]{dkhong@pusan.ac.kr} 
\affiliation{Department of
Physics,   Pusan National University,
             Busan 46241, Korea}

             \author{Jong-Wan Lee}
%\email[E-mail: ]{jwlee823@pusan.ac.kr} 
\affiliation{Department of
Physics,   Pusan National University,
             Busan 46241, Korea}

\author{Biagio Lucini}
\affiliation{College of Science, Swansea University, Singleton Park, Swansea SA2 8PP, UK}

\author{Maurizio Piai}
\affiliation{College of Science, Swansea University, Singleton Park, Swansea SA2 8PP, UK}

\author{Davide Vadacchino}
\affiliation{College of Science, Swansea University, Singleton Park, Swansea SA2 8PP, UK}

\date{\today}

\begin{abstract}
We conjecture that in Yang-Mills theories the ratio between the
ground-state glueball mass squared and the string tension is
proportional to the ratio of the eigenvalues of quadratic Casimir
operators in the adjoint and the fundamental representations. The
proportionality constant depends on the dimension of the space-time
only, and is henceforth universal. We argue that this universality,
which is supported by available lattice results, is a direct consequence
of area-law confinement. In order to explain this universal
behaviour, we provide three analytical arguments, based respectively on
a Bethe-Salpeter analysis, on the saturation of the scale anomaly by the
lightest scalar glueball and on QCD sum rules, commenting on the
underlying assumptions that they entail and on their physical implications. 
\end{abstract}
%\pacs{11.25.Tq,04.70.Bw,74.20.-z}
%%% 11.25.Tq Gauge/string duality
%%% 11.10.Kk Field theory in dimension other than four
%%% 11.25.Wx String and brane phenomenology
%%% 12.38.Cy Summation of perturbation theory in QCD
%%% 12.60.Fr Extension of electroweak Higgs sector
%%% 12.15.Ji Applications of electroweak models to specific processes
%%% 12.15.Lk Electroweak radiative corrections
%%% 12.60.Nz Technicolor models
%%% 13.40.Gp Electromagnetic form factors
%%% 14.20.Gk Baryon resonances with S=0

\maketitle

%\newpage

%%%%%%%%%%%%%%%%%%%%%%%%%%%%%%%%%%%%%%%%%%%%%%%%%%%%%%%%%%%%%%%%%%%%%%%%%%%%%%%%%%%%%%%%%
\section{Introduction}
Yang-Mills (YM) theories without matter fields are believed to exhibit a confining phase at low energies, in which
all bound states (glueballs) are gapped and color-singlet.
Confinement in YM theories is  supported by lattice studies~\cite{Greensite:2003bk}.
However, since glueballs are nonpertubative objects, we do not have yet good understanding of the properties of glueballs 
such as their mass spectrum or decay widths. 

It has been suggested that color confinement can be described in terms of  a dual Higgs 
mechanism or monopole condensation~\cite{tHooft:1977nqb,Mandelstam:1974pi,Seiberg:1994rs}. In this picture,  
monopoles, dual to color charges, condense in the color-confined phase, and  't Hooft operators 
develop a vacuum expectation value. The dynamical scale $\kappa$ is set by the condensate, 
which should be responsible for all other dimensional quantities in the confined phase. Monopole condensation implies 
a linear potential between a pair of static color charges, or equivalently an area law for the Wilson loop. 

In this letter we provide theoretical arguments and numerical evidence for the existence of a new universal law. The law states that 
the ratio of ground state glueball mass squared and the string tension is universally proportional to the ratio of the eigenvalues of quadratic Casimir operators for all confining gauge theories. 
The proportionality constant is independent of the gauge group and the strength of coupling as long as 
the area law arises. It depends only on the dimensionality of the space-time.   

%%%%%%%%%%%%%%%
\section{Glueball mass}

Calculating the ground state glueball mass is tantamount to showing that there is a gap in the ground state of pure YM theory, 
which has never been proved analytically except in three dimensions~\cite{Karabali:1998yq}. 
Numerical calculations of  glueball masses on the  lattice show the existence of a gap in YM theories~\cite{Morningstar:1999rf,Chen:2005mg,Lucini:2004my,Lucini:2012gg,Athenodorou:2016ebg,Athenodorou:2015nba,Lau:2015cna,Lau:2017aom}. 

Asymptotically, for  confining YM theories, the expectation value of rectangular Wilson loops
${\cal C}$ can be written as 
\begin{equation}
\left<W({\cal C})\right>=\left<\frac{1}{N}e^{i\oint_{\cal C}A}\right>=\exp\left[-\sigma LT+\cdots\right],
\label{area}
\end{equation}
where $LT$ is the area of  ${\cal C}$, $\sigma$ is the string tension between a static quark-antiquark pair,
and the ellipsis includes subleading corrections such as the L\"uscher term. 
Following the area law confinement, we write the string tension $\sigma$  as to define $\kappa$ via the proportionality to the quadratic Casimir operator on the 
fundamental representation
%\,\footnote{This is explicitly shown in 2+1 dimensions by constructing the vacuum wavefunction~\cite{Karabali:1998yq}.}
\begin{equation}
\sigma=\kappa^2\,C_2(F)\,,
\end{equation}
which is consistent with lattice results~\cite{Lucini:2001nv,Bali:2000un,Deldar:1999vi,Cardoso:2011cs,Lucini:2004my,Bringoltz:2008nd}.
The glueball is a bound state of adjoint gluons. On dimensional grounds,  its mass should be proportional to $\kappa$. 
For the ground state glueball we conjecture
\begin{equation}
m_{0^{++}}^2=\eta\,\kappa^2\,C_2(A)\,,
\end{equation}
where $\eta$ is  a universal ratio and $C_2(A)$  the quadratic Casimir for the adjoint representation.
The existence of the universal ratio $\eta$ is consistent with the large-$N$ universality of YM theories, 
supported by   Wilson loop calculations~\cite{Lovelace:1982hz} and 
gauge-gravity dualities~\cite{Imoto:2009bf}.
At finite $N$, the ratio of the eigenvalues of the relevant quadratic Casimir operators is~\cite{Slansky:1981yr} 
\begin{equation}
\frac{C_2(A)}{C_2(F)}=
\begin{cases} \frac{2N^2}{N^2-1}\;\,\text{for}\,\; \rm{SU}(N)\\
\frac{2(N-2)}{N-1}\;\,\text{for}\,\; \rm{SO}(N)\\
\frac{4(N+1)}{2N+1}\;\,\text{for}\,\; \rm{Sp}(2N)\,
\end{cases}\,,
\end{equation}
and  approaches 2 in the large-$N$ limit.

\begin{figure}[t]
\hspace{-.2in}
    \centering
    \begin{minipage}{0.48\textwidth}
        \centering
        \includegraphics[width=1\linewidth, height=0.25\textheight]{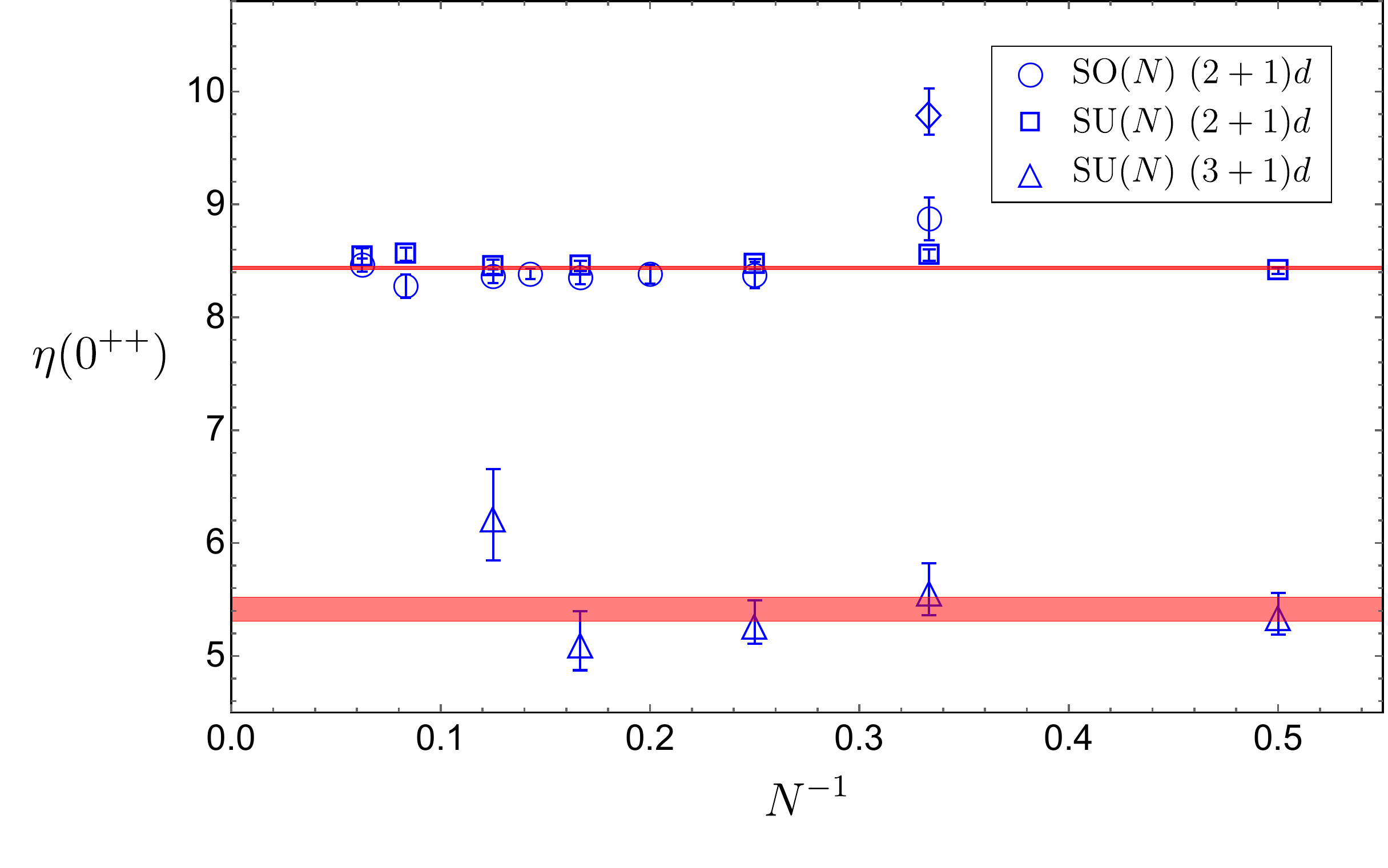}
      %  \caption{(a)}
       % \label{fig:prob1_6_2}
    \end{minipage}%
\hspace{.1in}
    \begin{minipage}{0.48\textwidth}
        \centering
        \includegraphics[width=1\linewidth, height=0.25\textheight]{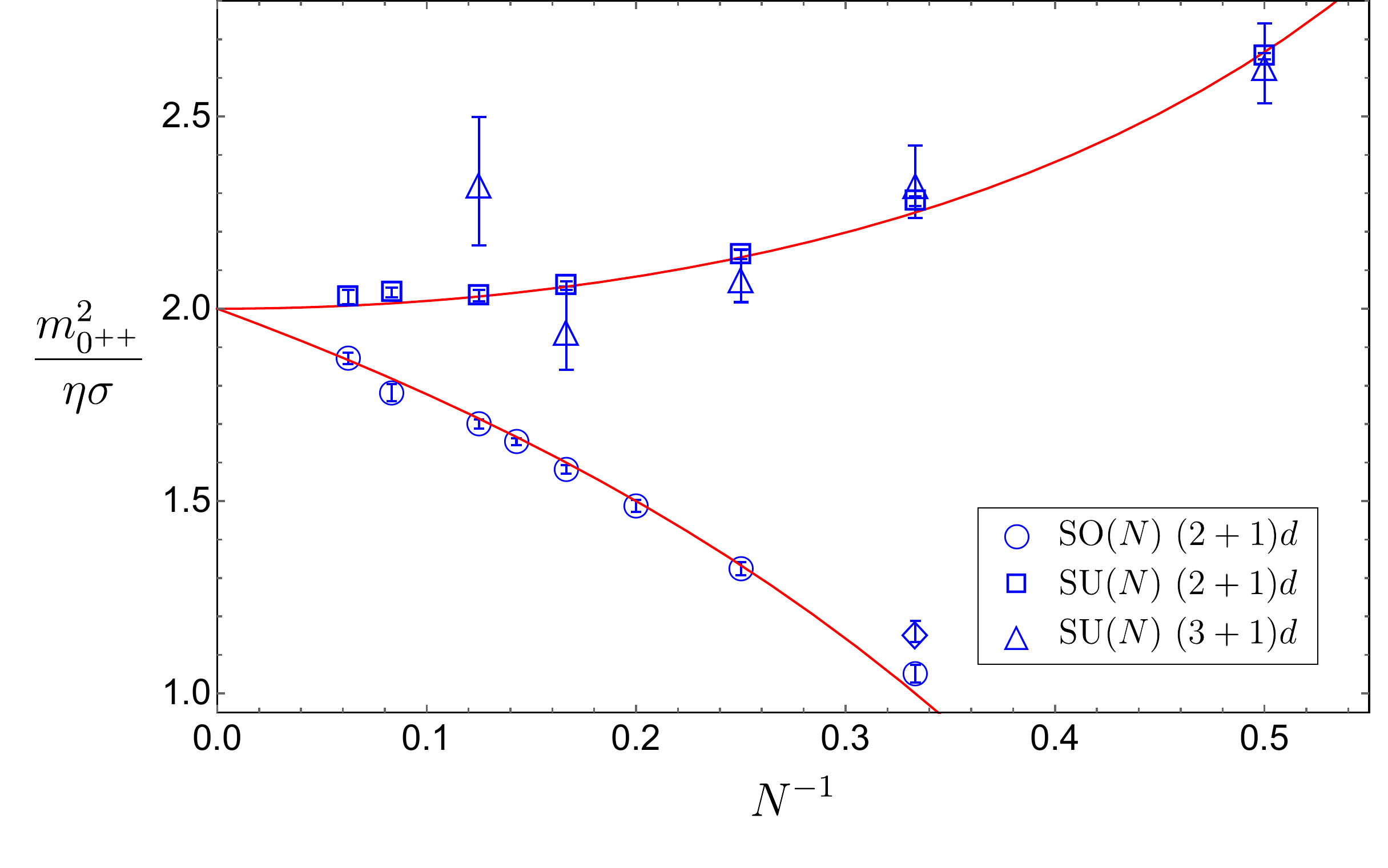}
     %   \caption{(b)}
       % \label{fig:prob1_6_1}
    \end{minipage}
  \caption{
The universal ratio $\eta$ (left panel), and glueball masses squared in units of the string tension (right panel),
 for various YM theories as a function of $1/N$.
The solid curves are the Casimir ratio $C_2(A)/C_2(F)$  for ${\rm SU}(N)$ (upper curve) and ${\rm SO}(N)$ (lower curve), respectively. The value of $\eta$ from  the tension of the ${\rm SO}(3)$ fundamental string is marked as $\diamond$.
}
  \label{fig1}   
\end{figure}

Glueball masses and string tensions have been calculated by various 
collaborations for YM theories in $3+1$ and $2+1$ dimensions~\cite{Morningstar:1999rf,Chen:2005mg,Lucini:2004my,Lucini:2012gg,Athenodorou:2016ebg,Athenodorou:2015nba,Lau:2015cna,Lau:2017aom}. 
% ${\rm SU}(N)$  results~\cite{Lucini:2004my,Lucini:2012gg,Athenodorou:2016ebg} 
%are compatible with  $1/N^2$ scaling of glueball masses in the large-$N$ limit.
%The $1/N$ scaling expected for ${\rm SO}(N)$ YM theories in 2+1 dimensions
% is also supported~\cite{Athenodorou:2015nba,Lau:2015cna,Lau:2017aom}.  
From the continuum-extrapolated lattice results of glueball mass and string tension, taking the data from the most recent large-$N$ calculations available in the literature~\cite{Lucini:2004my,Athenodorou:2015nba,Lau:2017aom}
(Fig.~\ref{fig1}), we find~\footnote{Our conjecture for the universal ratio is also supported by the analytic calculation of 
the ground-state glueball mass in 2+1 dimensional ${\rm SU}(N)$ gauge theories~\cite{Leigh:2006vg}, which finds $\eta(0^{++})\simeq8.41$, and suspected in the constituent gluon model in~\cite{Buisseret:2011bg}.}
\begin{equation}
\eta(0^{++})\equiv\frac{m_{0^{++}}^2}{\sigma}\cdot\frac{C_2(F)}{C_2(A)} =
\begin{cases} 5.41(12)\,,\;%\,\text{for}
(d=3+1)\,,\\
8.440(14)(76)\,,\;
%\text{for}\,\;
(d=2+1)\,.
\end{cases}
\label{eq_eta}
\end{equation}
%We took from the literature the  string tension and glueball masses 
%extracted from 
%the continuum extrapolation 
%and included  statistical errors only.
For  $3+1$ dimensions Eq.~(\ref{eq_eta})  is the constant fit of ${\rm SU}(N)$ results 
 over  $2\le N\le8$, with  $\chi^2/{\rm d.o.f.}\simeq 1$.
For  $2+1$ dimensions, lattice results  are available for ${\rm SU}(N)$, as well for ${\rm SO}(N)$,  with $2\le N\le16$,
hence we performed a constant fit for the universal ratio $\eta$ of both data sets.~\footnote{
The string tension can be defined also for ${\rm SO}(3)$ by considering distances of the order of the confinement scale. Yet, it is affected by large 
systematic uncertainties due to its instability~\cite{Athenodorou:2015nba,Lau:2017aom}. 
To mitigate the systematics, instead of this quantity, we use the string 
tension obtained from the fundamental of ${\rm SU}(2)$, assuming Casimir scaling for the 
string tension. 
We checked that by using the measured value of the string tension of ${\rm SO}(3)$, the value of $\eta$ 
does not change but yields a poor $\chi^2/{\rm d.o.f} \simeq 4.8$. 
%Resolving this issue would  require a dedicated numerical study.
% would certainly be 
%useful to further test the universal Casimir scaling discussed in this letter.
}
The resulting statistical error is quoted in the first parentheses in Eq.~(\ref{eq_eta}), 
with somewhat larger value of $\chi^2/{\rm d.o.f.}\simeq 1.9$.\,\footnote{The $\chi^2$ distribution does not improve significantly, even if the data for the lowest $N$ is excluded.} 
%This result could be understood as lattice data underestimate the uncertainties of the lattice measurement.

\begin{figure}[t]
\hspace{-.2in}
    \centering
    \begin{minipage}{0.48\textwidth}
        \centering
        \includegraphics[width=1\linewidth, height=0.25\textheight]{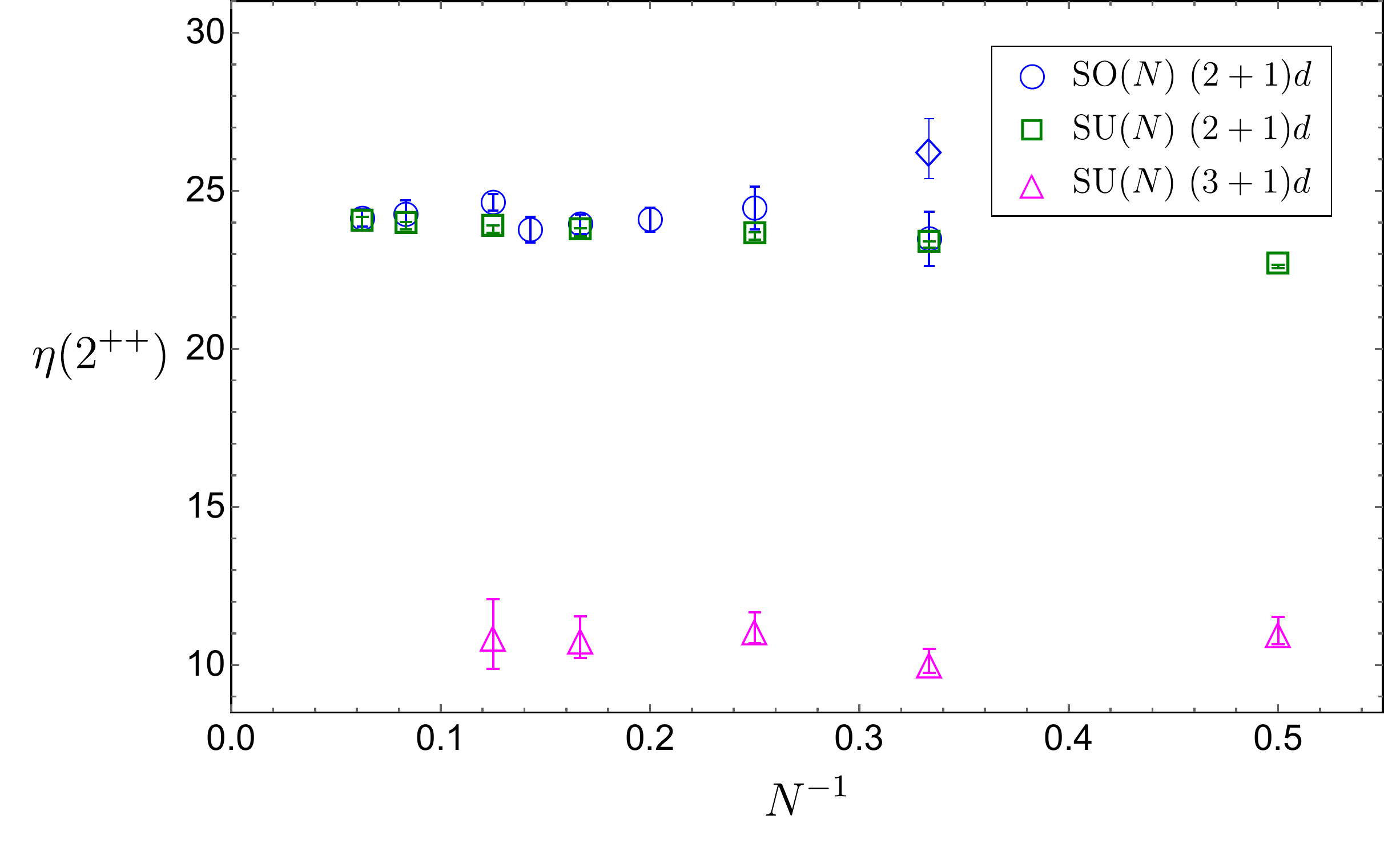}
      %  \caption{(a)}
       % \label{fig:prob1_6_2}
    \end{minipage}%
\hspace{.1in}
    \begin{minipage}{0.48\textwidth}
        \centering
        \includegraphics[width=1\linewidth, height=0.25\textheight]{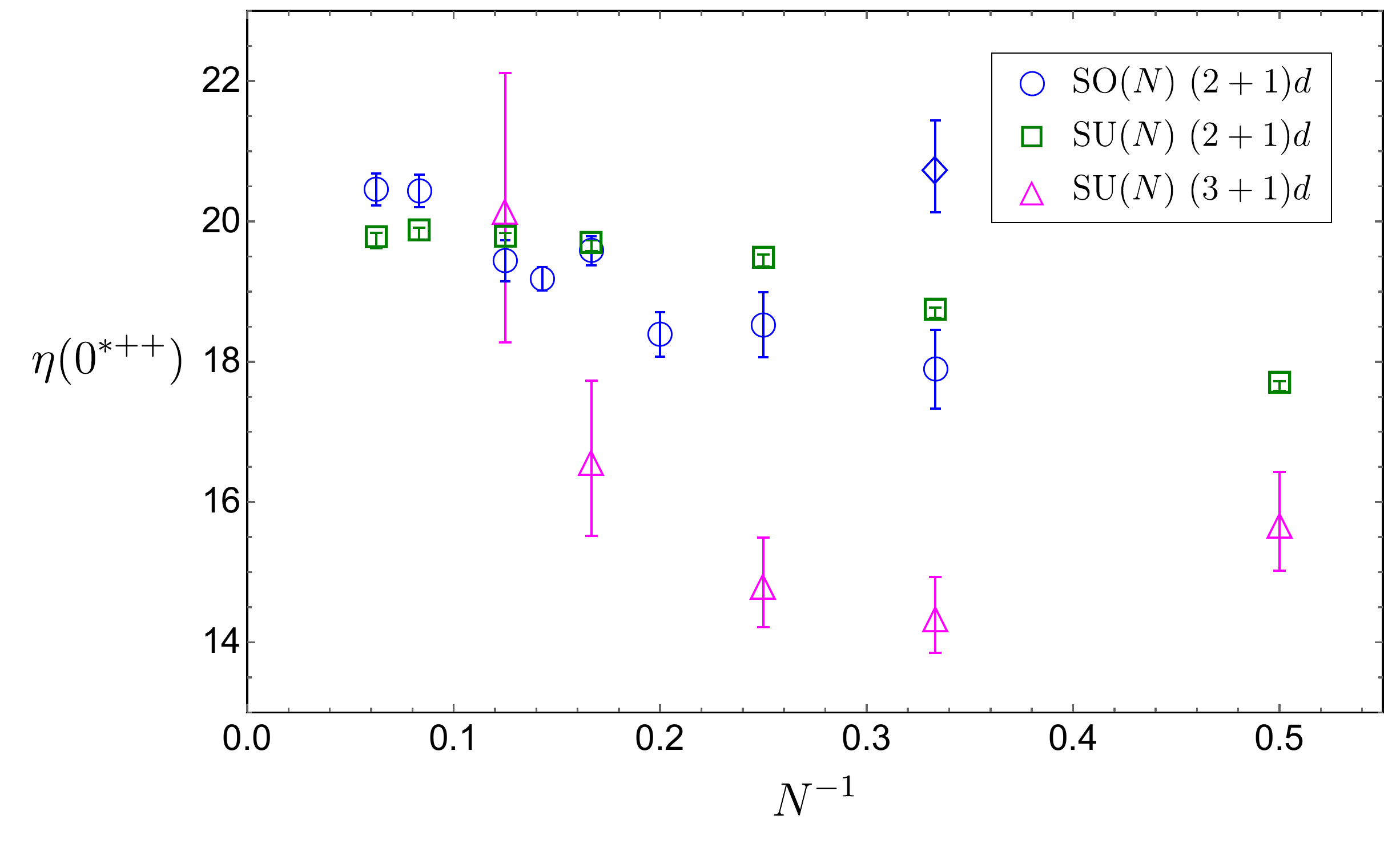}
     %   \caption{(b)}
       % \label{fig:prob1_6_1}
    \end{minipage}
  \caption{
The  ratio $\eta$ for the lowest-lying $2^{++}$ and $0^{\ast ++}$ (first excitation in the scalar channel) 
 as a function of $1/N$. The value of $\eta$ from  the tension of the ${\rm SO}(3)$ fundamental string is marked as $\diamond$.
}
  \label{fig2}   
\end{figure}

Deviations from universality in 2+1 dimensions between two classes of gauge groups
are assessed by calculating $\eta$ separately.
We find $\eta=8.386(25)$ ($\chi^2/{\rm d.o.f.}\simeq 1.3$) for ${\rm SO}(N)$ and $\eta=8.462(16)$ ($\chi^2/{\rm d.o.f.}\simeq 1.9$)
for ${\rm SU}(N)$.
%The discrepancy sits at the $1\%$ level. 
Given the expectation that the large-$N$ limit of the two sets should coincide, this difference of $3\sigma$ level is probably due to the systematic errors in the lattice data. We account for the discrepancy with a systematic error reported in the second parenthesis in Eq.~(\ref{eq_eta}).
%The result does not change significantly even if we exclude the small-$N$ 
%data. %, where the two sequences of theories should approach each other.
We also studied two heavier states, the $2^{++}$ glueball and the first excited scalar glueball, $0^{\ast++}$. 
The excited states start to see the deviation from the area-law confinement, hence it is not surprising that the $0^{\ast++}$ does not show universal behavior. (See Fig.~\ref{fig2}).  For the $2^{++}$, however, it is inclusive, because the constant fit gives a poor $\chi^2/{\rm d.o.f.}\simeq 19$ for the $2^{++}$  tensor glueballs in 2+1 dimensions, while it fits much better in 3+1 dimensions with $\chi^2/{\rm d.o.f.}\simeq1.1$.

%%%%%%%%%%%%%%%%%%%%%%%%%%%%%%%%%%%
\section{glueball mass and Casimir scaling}
Motivated by the strong numerical evidence for Casimir scaling, 
 we provide three analytical arguments to explain its origin.
None of the arguments is fully conclusive, as they all rely on specific dynamical assumptions that we highlight explicitly,
yet the picture that emerges is that Casimir scaling of ground state mass should capture much of the essence of the 
confinement properties of YM theories.

%%%%%%%%%%%%%%%%%%%%%%%%%%%%%%%%%%%%
\subsection{Bethe-Salpeter equation}
%{\it Bethe-Salpeter equations for glueballs}
The amplitude for creating two gluons out of vacuum to form a color-singlet bound state of momentum $P$ with a polarization $\lambda$
can be defined as 
\begin{equation}
\Gamma_R^{\mu\nu}(x_1,x_2;P,\lambda)=\left<0\right|{\rm T}A^{\mu\,a}(x_1)A^{\nu\,a}(x_2)\left|R(P,\lambda)\right>\,,
\end{equation}
where ${\rm T}$ denotes the time-ordered product and $\left<0\right|$ is the vacuum. Summation over 
color indices $a$ is understood.

\begin{figure}[h!]
  \includegraphics[width=3.5in]{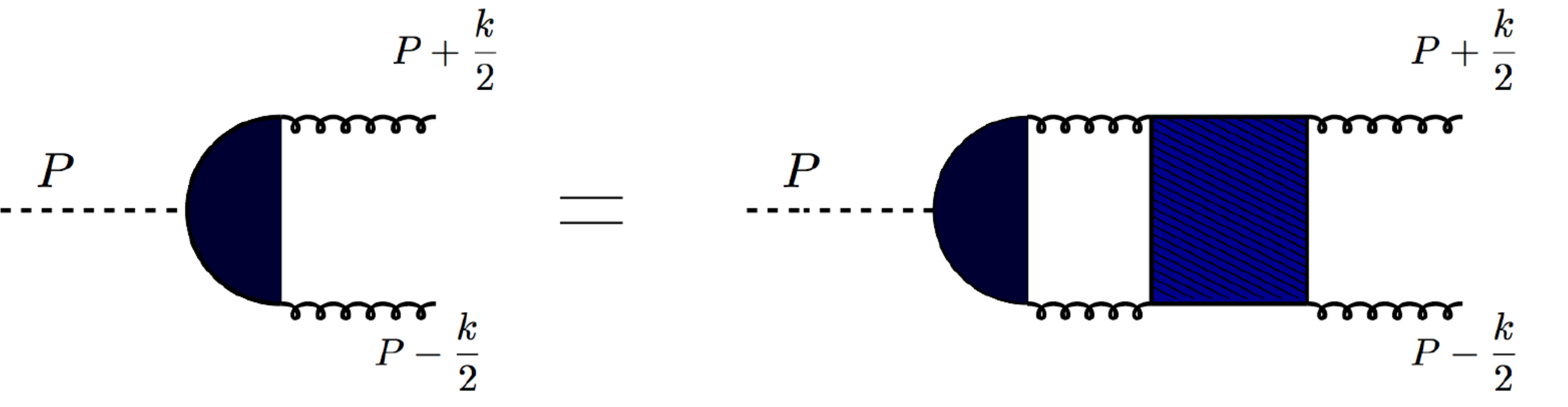}
  \caption{The BS equation for the glueballs. The half disk denotes the  BS amplitude of momentum $P$ and the relative momentum $k$ of two gluons. The box denotes the BS kernel.}
  \label{fig3}  
\end{figure}

The bound state amplitude satisfies the Bethe-Salpeter (BS) equations, 
obtained from the gluon four-point scattering amplitude near the pole, which 
are diagrammatically shown for the amputated BS amplitude in Fig~\ref{fig3}. 
% \vskip -1in

 From the BS equation,  the  scalar (amputated) amplitude $\chi_P$ obeys,
 in  Euclidean space, 
\begin{equation}
\left[\partial^2-P^2\right]\chi_P(x)=\int {\rm d}^4y\,V(x-y) \chi_P(y)\,,
\end{equation}
with $x=x_1-x_2$  the displacement of two external gluons. 

The area law for confinement is associated with the Regge behavior of the  spectrum:
$
M_{n}^2\sim n
$,
where $n=1,2,\cdots$ are the radial quantum numbers, 
reproduced by the approximate BS kernel
\begin{equation}
V(x-y)\approx\frac12\omega^2x^2\,\delta^4(x-y)\,.
\label{kernel}
\end{equation}
The BS kernel is nothing but the four-point function of gluons, properly projected for the spin-0 state.
If the string flux picture holds for the glueball states, $\omega$ should be the string tension of the Nambu-Goto action for the closed string that describes glueballs, $\omega\sim \sigma(A)=C_2(A)\kappa^2$ (see for example Eq. 2.26 in~\cite{Meyer:2004gx}). 
The radially excited scalar glueball mass is then  (for $n=1,2\cdots$)
\begin{equation}
M_{n}^2\sim\, C_2(A)\kappa^2\,(n+1)\,.
\end{equation}
Since the string tension is  $\sigma=\kappa^2\,C_2(F)$,  for the mass of ground state  $(n=1)$ glueball we find
\begin{equation}
\frac{m_{0^{++}}^2}{\sigma}=\eta\,\frac{C_2(A)}{C_2(F)}\,.
\label{casimir}
\end{equation}
%which is universal for all gauge groups that exhibit area-law confinement. 

There are corrections to Casimir scaling, coming from the corrections to the area law in Eq.~(\ref{area}). 
But such corrections are suppressed, arising at the next-to-next-to-leading order. 
Namely, the L\"uscher term in the expectation value of the Wilson loop in~(\ref{area}) does not modify  Casimir scaling, Eq.~(\ref{casimir}), 
since the L\"uscher term is a universal number~\cite{Luscher:1980fr} correcting the BS kernel by a shift itself
proportional to the Casimir; for $|x|\gg\kappa^{-1}$
\begin{equation}
\frac12\left( C_2(A)\kappa^2\,x-\frac{\alpha}{x}\right)^2\approx \frac12 C_2(A)^2\kappa^4x^2-\alpha\,C_2(A)\kappa^2\,,
\end{equation}
where $\alpha=(D-2)\pi/24$ is the universal coefficient of the L\"uscher term in $D$ dimensions. The ground state glueball mass then is corrected as
\begin{equation}
m_{0^{++}}^2\sim C_2(A)\kappa^2\,\left(2-\alpha\right)\,,
\end{equation}  
which does not change the universal scaling law.
% but the ratio $\eta$ is renormalized to $\eta\,(1-\alpha/2)$ due to the L\"uscher term. 

The corrections  become more important at high energies (short distances), in particular for the excited states,
for which we expect violations of the Casimir scaling to show.
%{\it These arguments rely upon are the factorization of the four-point function
%(exact at large-$N$) {\bf MP: is this true?}}, and the Regge behavior of the spectrum.
%The latter  is  affected for $n=1$ by the fact that 
As discussed in following, the characteristic behavior of the $0^{++} $ ground state may be understood in terms of its special role 
with respect to the scale symmetry of the system. 

%%%%%%%%%%%%%%%%%%%%%%%%%%%%%%%%%%%
\subsection{Scale anomaly}
Pure Yang-Mills theories in four dimensions are classically scale invariant, but scale symmetry is anomalous, broken by quantum effects. 
Futhermore, the (anomalous) scale symmetry in YM theory is spontaneously broken as well, since the YM vacuum develops a non-vanishing expectation value for the order parameter for confinement.  If the scale anomaly is parametrically  small, compared to such vacuum expectation value of the order parameter, then there should be a (pseudo) Nambu-Goldstone boson in YM theory, associated with spontaneous breaking of the scale symmetry. 
Namely, by Goldstone's theorem, the dilatation current, associated with the scale symmetry $x^{\rho}\to e^{\lambda}x^{\rho}$, creates a state, called dilaton:
\begin{equation}
\left<0\right|D^{\mu}(x)\left|D(p)\right>=-if_D\,p^{\mu}e^{-ip\cdot x}\,,
\label{dilaton}
\end{equation}
where the dilatation current $D_{\mu}=x^{\nu}\theta_{\mu\nu}$ with the improved energy-momentum tensor $\theta_{\mu\nu}$~\cite{Callan:1970ze} and $f_D$ is the dilaton decay constant. 

The scale anomaly in pure YM theory is given as 
\begin{equation}
\partial^{\mu}D_{\mu}=-\frac{\beta(g)}{2g}\,F^a_{\mu\nu}F^{a\,\mu\nu}\,,
\label{anomaly}
\end{equation}
where $\beta(g)$ is the beta function  and $F_{\mu\nu}^a$ the field-strength tensor. Since the divergence of the dilatation current can be written in terms of the trace of the energy-momentum tensor
as $\partial_{\mu}D^{\mu}=\theta^{\mu}_{\mu}$, the anomalous Ward identity~(\ref{anomaly}) relates the two-point function of the trace of the energy-momentum tensor to its one-point function or the scale anomaly:
\begin{equation}
\int_x\left<0\right|{\rm T}\,\theta^{\mu}_{\mu}(x)\theta^{\nu}_{\nu}(y)\left|0\right>=-4\left<\theta^{\mu}_{\mu}(y)\right>\,.
\label{wi}
\end{equation}
As all the gluons equally and additively contribute to the vacuum energy, the scale anomaly should be given on dimensional ground as $\left<\theta^{\mu}_{\mu}\right>=-\tilde\beta\,C_2(A)\kappa^4$, after subtracting out the part that is independent of the condensate.  If the scale anomaly is parametrically small or $|\tilde\beta|\ll1$, there should be a light dilaton, defined as in Eq.~(\ref{dilaton}), that saturates the two-point function in~(\ref{wi}):
\begin{equation}
\int_x \left<0\right|{\rm T}\,\theta^{\mu}_{\mu}(x)\theta^{\nu}_{\nu}(y)\left|0\right>\approx \!
%\iint_x \frac{{\rm d}^3p}{(2\pi)^3}\left<0\right|\theta^{\mu}_{\mu}(x)\left|D(p)\right>\frac{i}{p^2-m_D^2}\left<D(p)\right|\theta^{\nu}_{\nu}(y)\left|0\right>=
f_D^2m_D^2\,.
\label{pcdc}
\end{equation}
We then have a so-called partially conserved dilatation current (PCDC) relation, 
\begin{equation}
f_D^2m_D^2=-4 \left<\partial_{\mu}D^{\mu}\right>=-16\,{\cal E}_{\rm vac}\,,
\label{scale anomaly}
\end{equation}
where ${\cal E}_{\rm vac}=-\tilde\beta\,C_2(A)\kappa^4/4$ is the vacuum energy density of YM theories in the confined phase.
The vacuum energy (density) scales as $C_2(A)$.  On the other hand the dilaton decay constant, $f_D$, 
measures the strength of the amplitude that creates the dilaton out of vacuum, 
which should not depend on the number of gluon fields, but only on the characteristic scale $\kappa$ that defines the scale of spontaneous scale-symmetry breaking. 
We hence find the ground state glueball mass $m_{0^{++}}^2\propto \tilde\beta\, C_2(A)\kappa^2$, if identified as dilaton, and it becomes parametrically small if $\tilde\beta\ll1$.

The assumption about saturation of the Ward identity by the lightest $0^{++}$ state, Eq.~(\ref{pcdc}),
is equivalent to assuming  the existence of a weakly-coupled low-energy effective field theory for 
the $0^{++}$ state in terms of the dilaton field, in spite of its mass not being particularly small, compared to other excited states like $0^{\ast++}$ glueball states, which implies $\tilde\beta\sim1$. The fact that 
Casimir scaling holds for the ground state glueball but not for excited states (as hinted also by lattice calculations) is therefore quite intriguing, and very distinctive from analysis based on other approaches that do not differentiate the lightest state. 

 %This seems to suggest that the ground state glueball does behave like dilaton and the excited glueball states have a significantly smaller overlap with the trace of the energy-momentum tensor. 

%%%%%%%%%%%%%%%%%%%%%%%%%%%%%%%%%
\subsection{Sum rules}
The glueball mass can be extracted from the correlators of  interpolating operators made of gluons. 
For scalar glueballs one considers the correlator of the gluonic field strength tensor
${\cal O}_S(x)\equiv \alpha_sF_{\mu\nu}^aF^{a\,\mu\nu}$: 
\begin{equation}
\Pi_{S}(x)=\left<0\right|{\rm T}\left[{\cal O}_S(x){\cal O}_S(0)\right]\left|0\right>=\sum_n\,c_n\,e^{-m_n\left|x\right|}\,,
\label{cor}
\end{equation} 
where ${\rm T}$ is the time-ordering operation and the smallest $m_n$ will be the mass of ground state glueball $0^{++}$.
 
The sum rules, associated with the moments of the correlators, exploit the operator production expansion.
For the zero moment, one finds~\cite{Novikov:1980dj}
\begin{equation}
\int{\rm d}^4x\,\Pi_S(x)=\frac{32\pi^2}{b}\left<0\right|\frac{\alpha_s}{\pi}F_{\mu\nu}^aF^{a\,\mu\nu}\left|0\right>\,,
\end{equation}
where $b$ is the first coefficient of the beta function and the integral is regularized by subtracting out the perturbative contributions.
Assuming 
single-particle states to be stable and
inserting a complete set between the interpolating operators in (\ref{cor}), we have
\begin{equation}
\sum_{n=0}^{\infty}\,f_n^2\,m_n^2=\frac{32\pi^2}{b}\left<0\right|\frac{\alpha_s}{\pi}F_{\mu\nu}^aF^{a\,\mu\nu}\left|0\right>\,,
\label{sr}
\end{equation} 
where the decay constants $f_n$ are normalized by
\begin{equation}
f_n\,m_n^2\equiv \left<0\right|{\cal O}_S(p)\left|n\right>_{p^2=0}\,.
\end{equation}
Because of the summation over gluons in the condensate in (\ref{sr}), 
we expect the scalar glueball mass squared to be proportional to  $C_2(A)$. 
We note the similarity with  the low energy theorem~(\ref{scale anomaly}), if the sum rule (\ref{sr}) is saturated by the ground state or, equivalently, for the excited states $f_nm_n^2\ll f_0m_0^2$, which suggests that the excited states have very little overlap with the operator ${\cal O}_S$. The numerical analysis we report in this letter seems to suggest that this is the case, as we do not see evidence of Casimir scaling in the excited states, but only in the ground state.

%%%%%%%%%%%%%%%%%%%%%%%%%%%%%%%%
\section{Discussions and conclusion}
For Yang-Mills theories, % based on any non-abelian group, 
we conjectured that the ground state 
glueball mass squared, measured in units of string tension, is 
universally proportional to the ratio of the eigenvalues of the quadratic Casimir operator of the adjoint over that of the fundamental representation. 
The conjecture relies on the  %existence of an 
area law for confinement,
and the specific coefficient should depend only on the dimensionality of the space-time,
but not on the specific group.

We provided three  analytical arguments to justify Casimir scaling, based respectively on the
Bethe-Salpeter equation, scale anomaly, and sum rules. 
We tested this law on existing numerical  lattice results in pure ${\rm SU}(N)$ and ${\rm SO}(N)$ 
Yang-Mills theories in $3+1$ and $2+1$ dimensions. 
The data strongly support  Casimir scaling for the ground state.  
The values of the universal constant extracted from  lattice data are $\eta(0^{++})=5.41(12)$ for 3+1 
dimensions and $\eta(0^{++})=8.444(15)(85)$ for 2+1 dimensions. 
Numerical results are 
inconclusive for the $2^{++}$ state, while showing that universality does not hold 
for the first excitation in the  $0^{ ++}$ channel.

If the conjectured universal scaling is confirmed, it would
shed light on the mechanism yielding confinement in YM theories. 
It would be therefore quite interesting  to test further numerically our conjecture
for other gauge groups such as $Sp(2N)$ and ${SO}(N)$ in $3+1$ dimensions,
and to extend the arguments discussed here to provide systematic control over
sub-leading corrections (if they exist) to exact Casimir scaling.
 %{\bf MP:  something more about the large-N expansion?}

%%%%%%%%%%%%%%%%%%%%%%%%%%%%%%%%%%%%%%%%%%%%%%%%%%%%%%%%%%%%%%%%%%%%%
%%%%%%%%%%%%%%%%%%%%%%%%%%%%%%%%%%%%%%%%%%%%%%%%%%%%%%%%%%%%%%%%%%%%%%%%%%%%%%%%%%%%%%%%%

%\vfill \eject

\acknowledgments
DKH thanks L. Giusti and C. Kim for useful comments and the CERN theory group for the hospitality during his visit. We thank  A. Armoni, G. Aarts and F. Buisseret for comments. 
This research was supported by Basic Science Research Program through the National Research Foundation of Korea (NRF) funded by the Ministry of Education (NRF-2017R1D1A1B06033701) (DKH) and also by Korea Research Fellowship program funded by the Ministry of Science, ICT and Future Planning through the National Research Foundation of Korea (2016H1D3A1909283) and under the framework of international  cooperation program  (NRF-2016K2A9A1A01952069) (DKH and JWL).
The work of BL, MP and DV is supported in part  by the STFC grant ST/L000369/1.

\end{document}